\documentclass[12pt]{article}
\usepackage{epsfig}

\usepackage{amssymb}
\usepackage{amsmath}
\usepackage{amsfonts}
\usepackage{amsthm}

\newcommand{\bbr}{\bra\!\bra}

\newcommand{\kt}{\rangle}
\newcommand{\bra}{\langle}

\def\be{\begin{equation}}
\def\ee{\end{equation}}
\def\ba{\begin{array}{c}}
\def\ea{\end{array}}
\def\p{\partial}
\def\ben{$$}
\def\een{$$}

 \newtheorem{thm}{Theorem}[section]
 \newtheorem{cor}[thm]{Corollary}

 \newtheorem{conj}[thm]{Conjecture}
 \theoremstyle{definition}
 
 \theoremstyle{remark}

\begin{document}

\begin{center}

{\Large \bf PT-symmetric quantum models living in an auxiliary
Pontryagin space}

\vspace{1cm}

 \vspace{1cm}

{Miloslav Znojil}

\vspace{1cm}

 Nuclear Physics Institute ASCR

 250 68 \v{R}e\v{z}, Czech Republic

\vspace{1cm}

email: znojil@ujf.cas.cz

\end{center}

\vspace{.13cm}


\vspace{1cm}

KEYWORDS

\vspace{.16cm}

 \noindent
{quantum mechanics, Hermitizations of observables, auxiliary Krein
and Pontryagin spaces, Jacobi-matrix Hamiltonians, Dieudonn\'{e}
equation}

\vspace{1cm}

\section*{Abstract}

The recent heuristic as well as phenomenological success of the use
of non-Hermitian Hamiltonians which are required self-adjoint in a
Krein space ${\bf K}$ is recalled, and an extension of the scope of
such a version of quantum theory is proposed. The usual choice of
the indefinite metric ${\cal P}$ treated as the operator of parity
is generalized. {\it In nuce}, the operators ${\cal P}$ are admitted
to represent the indefinite metric in a Pontryagin space $\tilde{\bf
K}$. A constructive version of such a generalized quantization
strategy is outlined and found feasible.

\vspace{1cm}

\section{Introduction}

In the most common applications of quantum theory \cite{Messiah} the
norm-preserving time-evolution of a non-relativistic quantum system
is controlled by a self-adjoint Hamiltonian
$\hat{H}=\hat{H}^\dagger$. One starts from its form defined,
typically, as a sum of kinetic energy $-\triangle$ and of an
interaction-energy potential $V({x})$. The resulting operator is
assumed acting in a friendly Hilbert space ${\bf H}^{(F)}$
represented, say, by the linear space $L^2(\mathbb{R})$ of
quadratically integrable functions. In the Schr\"{o}dinger's mode of
description the states are  ket vectors $|\psi(t)\kt$ living in
${\bf H}^{(F)}$. Their form may be determined via Schr\"{o}dinger
equation
 \be
 {\rm i}\p_t \,|\psi(t)\kt=\hat{H}\,|\psi(t)\kt\,,
 \ \ \ \ \ |\psi(t)\kt \in {\bf H}^{(F)}
 \,.
 \label{SEtD}
 \ee
In constructive considerations, unitary Fourier-like transformations
$\Omega$ of ${\bf H}^{(F)}$ are often used leading to the strictly
equivalent physical predictions. Once the Fourier-like maps $\Omega$
are allowed more general and non-unitary, the images
$\Omega\,|\psi(t)\kt^{(F)} :=|\psi(t)\kt^{(P)}$ of the kets
$|\psi(t)\kt^{(F)} \in {\bf H}^{(F)}$ should be treated as
transferred into a {\em non-equivalent} Hilbert space, $
|\psi(t)\kt^{(P)} \in {\bf H}^{(P)}$ \cite{SIGMA}.

Whenever the ``friendly" space ${\bf H}^{(F)}$ is assumed
unphysical, we are well motivated to treat the other space ${\bf
H}^{(P)}$ as correct and ``physical". If the new space ${\bf
H}^{(P)}$ remains friendly, we may just change the Hamiltonian
operator accordingly,
 \be
 \Omega: H \to \mathfrak{h} := \Omega\,H\,\Omega^{-1}\,.
 \label{change}
 \ee
The truly nontrivial situation only emerges when the new,
self-adjoint Hamiltonian $\mathfrak{h}$ defined in ${\bf H}^{(P)}$
becomes prohibitively complicated. Then, the pullback of the
Hermiticity condition to the friendly space is still recommendable
yielding the hidden Hermiticity rule (a.k.a. ``crypto-Hermiticity"
or ``quasi-Hermiticity")
 \be
 H^\dagger\Theta=\Theta\,H\,,\ \ \ \ \
 \Theta=\Omega^\dagger\Omega\,
 \label{dieudonne}
 \ee
postulated directly in the friendly space ${\bf H}^{(F)}$ and,
historically, attributable to Dieudonn\'e \cite{Dieudonne}.

In the present paper we intend to pay attention to the details of
the path between the ``false" initial Hilbert space  ${\bf H}^{(F)}$
and its concrete, correct and ``physical" amendment ${\bf H}^{(P)}$.
We shall analyze and generalize one of the most popular strategies
of transition ${\bf H}^{(F)}\to {\bf H}^{(P)}$ which makes a detour
via an intermediate auxiliary Krein space ${\bf K}$ (based on the
use of an indefinite (pseudo)metric ${\cal P}$) and which might be
called ${\cal PT}-$symmetric quantum mechanics (PTSQM, \cite{Carl}
-- more comments and explanations will be added below).


\section{The quantization recipe based on the ${\cal PT}$ symmetry \label{aIs}}

Our interest in the possibility of an amendment of the PTSQM recipe
was inspired by the particular success of the removal of the
ambiguity of $\Theta_\alpha(H)$ based on the {\em ad hoc} PTSQM
assumption that at $\alpha=\alpha_{exceptional}$, the product ${\cal
P}\Theta_\alpha(H)$ might preserve certain mathematical properties
of the parity \cite{pseudo} or, alternatively, that it could acquire
certain phenomenological features of the charge \cite{BBJ}.

In both of the latter scenarios, the most natural mathematical
interpretation of the operator ${\cal P}$ may be seen in its role of
a Krein-space (pseudo)metric. On such a background, the core of our
present main proposal will lie in the replacement of the
intermediate Krein space ${\bf K}$ by the alternative intermediate
Pontryagin space $\tilde{\bf K}$, dictated by the intention of
making the auxiliary pseudometric operator ${\cal P}$ much less
dependent upon the phenomenological notion of the observable parity.

Our present second guiding idea is that due to the isospectrality of
$H$, $H^\dagger$ and $\mathfrak{h}$ one often finds it useful to
stay working inside ${\bf H}^{(F)}$. A persuasive illustration of
such a three-Hamiltonian strategy and scenario has been offered,
almost twenty years ago, by nuclear physicists \cite{Geyer}.
Nevertheless, from the purely practical and heuristic point of view,
the ultimate and decisive amendment of the recipe only appeared in
the context of field theory \cite{BG,Kim}. The most productive trick
has been found in an {\em additional} postulate
 \be
 H^\dagger{\cal P} ={\cal P}\,H\,.
 \label{PT}
 \ee
The latter property of the Hamiltonian (where the symbol ${\cal P}$
denotes, most often, the operator of parity) is called its ${\cal
PT}-$symmetry (cf., e.g., reviews \cite{Carl,ali} for an explanation
of this terminological convention).

We should emphasize that in the major part of the the recent
literature on ${\cal PT}-$symmetry in physics the phenomenological
Hamiltonian is assumed given as a non-self-adjoint operator
$\hat{H}\neq \hat{H}^\dagger$ acting in an unphysical Hilbert space
${\bf H}^{(F)}$ {\em and} exhibiting the additional ``symmetry"
(\ref{PT}). As long as the underlying class of the admissible
non-unitary Hermitization mappings $\Omega$ is only very weakly
restricted in such a case, one of the main weak points of the theory
may be seen in the ambiguity of the assignment $H \to\mathfrak{h} $
given by Eq.~(\ref{change}), i.e., in the ambiguity of the choice of
the ``metric" operator out of a family $\Theta=\Theta_\alpha(H)$
where $\alpha=1,2,\ldots$ \cite{Geyer}.

The assumption of simplicity of the (pseudo)metrics ${\cal P}$ and
the additional natural assumption of its involutivity ${\cal P}^2=I$
usually decisively facilitate the construction of the physical
Hilbert space ${\bf H}^{(P)}$. In some considerations, it makes
sense to treat the Hilbert space ${\bf H}^{(P)}$ as a single element
of the whole family of the mutually unitarily equivalent spaces
among which a ``special" one will be denoted by the symbol ${\bf
H}^{(S)}$ -- according to Ref.~\cite{SIGMA} its superscript $^{(S)}$
might mean a ``synthesis" or ``sophistication".




In our compact review paper \cite{SIGMA} we proposed that the
relation between the equivalent representations ${\bf H}^{(P)}$ and
${\bf H}^{(S)}$ of the physical Hilbert space of states might be
visualized as an equivalence in which one works with the respective
generalized inner products $\bra \psi| \phi \kt \to \bra \psi| \phi
\kt^{(P,S)} :=\bra \psi| \Theta^{(P,S)}|\phi \kt$ using the concept
of the {\em ad hoc} metric operators such that $\Theta^{(S)}\neq
\Theta^{(P)} \equiv I$. In other words, one can speak about a
metric-dependent definition of the conjugation in ${\bf H}^{(S)}$.
In this manner one updates the usual, ``friendly" (sometimes called
``Dirac's") Hermitian conjugation of vectors,
 \be
 {\bf T}^{(F)}: |\psi(t)\kt \ \to \ \bra
 \psi(t)|
 \label{equaDi}
 \ee
which is active in the auxiliary, unphysical Hilbert space ${\bf
H}^{(F)}$. In the update one replaces it by the metric-dependent
prescription
 \be
 {\bf T}^{(S)}: |\psi(t)\kt \ \to \ \bbr
 \psi(t)| := \bra
 \psi(t)|\,\Theta\,.
 \label{equagg}
 \ee
The latter recipe should be read as active in the sophisticated
physical Hilbert space ${\bf H}^{(S)}$, the kets of which coincide
with those of ${\bf H}^{(F)}$. For our present purposes the
prescription (\ref{equagg}) may be identified, therefore, with the
traditional Hermitian conjugation used in the unusual,
``sophisticated" space ${\bf H}^{(S)}$.

The mathematics which is hidden behind the transition from
Eq.~(\ref{equaDi}) to Eq.~(\ref{equagg}) is fairly nontrivial. For
this reason (and also for the sake of brevity of our forthcoming
considerations) let us circumvent, in the present paper, the
majority of the technical subtleties connected with the underlying
functional analysis and let us restrict our attention just to the
finite-dimensional vector-space versions of the triplet of the
Hilbert spaces ${\bf H}^{(F,S,P)}$ in question.

\section{Illustrative Jacobi-matrix  toy-model Hamiltonians}

As we already mentioned, one of the most important emerging
questions (formulated and answered, in the context of physics, by
Scholtz et al \cite{Geyer}) concerns the ambiguity and/or
possibility of an identification of an ``optimal" Hilbert space
${\bf H}^{(S)}$ for a given Hamiltonian $\hat{H}$ with the real
spectrum. In the brief summary of this point let us employ the
notation of review \cite{SIGMA} where we suggested to write the
``hidden" Hermiticity condition in ${\bf H}^{(S)}$ in the following
abbreviated form
 \be
 \hat{H}=
 \hat{H}^\ddagger:=\Theta^{-1}
 \hat{H}^\dagger\,\Theta\,.
 \label{cryptoh}
 \ee
The superscript $^\dagger$ stands here for the ``Dirac's"
transposition plus complex conjugation as defined by
Eq.~(\ref{equaDi}) for vectors and as used, in the $N \to \infty$
limit, in the most common auxiliary-space representations ${\bf
H}^{(F)}=\ell^2(\mathbb{Z})$ or ${\bf
H}^{(F)}=\mathbb{L}^2(\mathbb{R})$ with $\Theta^{(F)}\,\equiv\,I$.
In this notation the ``doubled superscript" $^\ddagger$ marks the
(crypto)hermitian conjugation of Eq.~(\ref{cryptoh}) for the
operators in ${\bf H}^{(S)}$. In this manner the cryptohermitian
conjugation~(\ref{equagg}) of vectors is extended to the
cryptohermitian conjugation of operators in ${\bf H}^{(S)}$. Both
these relations contain the same nontrivial metric operator
$\Omega^\dagger\,\Omega =\Theta=\Theta^{(S)} \neq I$.

\subsection{Multiparametric chain models exhibiting an up-down symmetry}

In the physics literature as reviewed briefly in Ref.~\cite{Gegenb}
we witness an intensification of interest in the real and
$N-$dimensional tridiagonal-matrix Hamiltonians. The main reason is
that these  Hamiltonians $\hat{H}^{(N)}$ describe a rather universal
$N-$site quantum-lattice dynamics in which just the nearest-neighbor
interaction is taken into account. The second reason is that these
models are nontrivial in the sense that the real matrix
$\hat{H}^{(N)}$ itself (possessing, presumably, real and
non-degenerate spectrum) may remain asymmetric, i.e., manifestly
non-Hermitian in the linear-algebraic sense, $\hat{H}^{(N)}\neq
\left [\hat{H}^{(N)}\right ]^\dagger$.

In the language of physics the role of the latter Hamiltonians
(which cannot generate the unitary evolution inside the most common
real vector space ${\bf H}^{(F)}$) may be seen in their intimate
connection with experiments \cite{Makris}. At the same time their
mathematical analysis may significantly be simplified via a specific
choice of the matrix elements in $\hat{H}^{(N)}$. In
Refs.~\cite{horizons}, for example, we assumed that Hamiltonian
$\hat{H}^{(N)}$ representing a finite-dimensional
anharmonic-oscillator-like model is a diagonal matrix (with an
equidistant unperturbed spectrum) which is complemented by a small
antisymmetric term mimicking the nearest-neighbor interaction of a
chain-model type.

We revealed  that an enormous simplification of the analysis appears
when one adds another requirement of a parity-type symmetry of the
(real) matrix with respect to its second diagonal,
  \be
 H^{(chain)}
 =\left [\begin {array}{cccccc}
  1-N&g_1&0&0&\ldots&0\\
 -g_1& 3-N&g_{2}&0&\ldots&0\\
 0&-g_{2}&5-N&\ddots&\ddots&\vdots
 \\
 0&0&\ddots&\ddots&g_{2}&0
 \\
 \vdots&\vdots&\ddots&-g_{2}&N-3&g_{1}\\
 0&0&\ldots&0&-g_{1}&N-1
 \end {array}\right ]\,\neq \,\left (H^{(chain)}\right )^\dagger\,.
 \label{NNIPTS}
 \ee
In spite of the presence of many independent coupling constants such
a symmetry proved sufficient to guarantee  the reality of the
spectrum even far beyond the weak-coupling dynamical regime. Thus,
the ``hidden" Hermiticity materializes via the transition to the
``second physical" Hilbert space ${\bf H}^{(S)}$ in a fairly large
and non-numerically defined cryptohermiticity domain
$D=D(g_1,g_2,\ldots,g_J)$ of as many as $J=entier[N/2]$
independently variable couplings.

The pragmatic appeal of multiparametric models (\ref{NNIPTS}) has
been weakened by the purely numerical nature of the eligible metrics
$\Theta^{(N)}$ defining the alternative Hilbert spaces ${\bf
H}^{(S)}$. In the subsequent, more constructive
studies~\cite{fund,JMP} and ~\cite{Tater} we diminished our
phenomenological ambitions, therefore. We turned attention to the
more elementary, square-well-type discrete Hamiltonian matrices
endowed with the mere one-parametric point-like
Hermiticity-violating interaction terms located either near the
center or near the boundary walls, respectively. For example, the
option of Ref.~\cite{Tater} with
 \be
  H^{(N)}({\lambda})=  \left[ \begin {array}{cccccc}
 2&-1-{\it {\lambda}}&0&\ldots&0&0
\\
{}-1+{\it {\lambda}}&2&-1&0&\ldots&0
\\
{}0&-1&\ \ \ 2\ \ \ &\ddots&\ddots&\vdots
\\
{}\vdots&0&\ddots&\ \ \ \ddots\ \ \ &-1&0
\\
{}0&\vdots&\ddots&-1&2&- 1-{\it {\lambda}}
\\
{}0&0&\ldots&0&-1+{\it {\lambda}}&2
\end {array}
 \right]\,
 \label{toym}
 \ee
enabled us to reach a certain next-to-solvable status of
transparency of transitions between the unphysical and physical
Hilbert spaces ${\bf H}^{(F)}$ and ${\bf H}^{(S)}$, respectively.
For the latter family of non-unique candidates for the
physics-representing Hilbert spaces we were able to offer a
manifestly constructive and {\em complete} classification of the
admissible metrics $\Theta=\Theta(H)$.

\subsection{Asymmetric solvable models}

A residual weakness of the latter model (\ref{toym}) has been felt
in a comparatively complicated structure of its bound-state
eigenvectors. This observation motivated our subsequent search for
another one-parametric toy-model Hamiltonian characterized by a
non-numerical solvability of Schr\"{o}dinger equation
 \be
  \hat{H}^{(N)}\,|\psi_n\kt=E_n\,|\psi_n\kt\,,\
   \ \ \ n = 0, 1, \ldots\,N-1\,
 \label{SEb}
 \ee
in ${\bf H}^{(F)}$ at any matrix dimension $N=1,2,\ldots$. A very
special Hamiltonian of the required type, viz.,
 \be
  \hat{H}^{(N)}:= H^{(N)}(a)=\left[
\begin {array}{ccccc}
a+1&-1&0&0&\ldots\\
{}-a-1&a+3&-2&0&\ldots\\
{}0&-a-2&a+5&-3&\ddots\\
{}0&0&-a-3&a+7&\ddots\\
{}\vdots&\vdots&\ddots&\ddots&\ddots
\end {array} \right]\,
 \label{wodel}
 \ee
has been proposed in Ref.~\cite{Gegenb}. Such a Hamiltonian must be
truncated. At finite cut-offs $N<\infty$ it offers an interesting
phenomenological model admitting the non-numerical diagonalization
as well as a systematic construction of the complete set of the
eligible metrics $\Theta$ via Dieudonn\'{e} equation. Unfortunately,
the loss of the up-down symmetry in model (\ref{wodel}) proved to
lead to a loss of the nice properties in the limit $N \to \infty$.
For this reason we finally decided to turn our attention to a
compromising asymmetric version of model (\ref{toym}) here,
 \be
  H^{(N)}({\lambda})=  \left[ \begin {array}{cccccc}
 2&-1-{\it {\lambda}}&0&\ldots&0&0
\\
{}-1+{\it {\lambda}}&2&-1&0&\ldots&0
\\
{}0&-1&\ \ \ 2\ \ \ &\ddots&\ddots&\vdots
\\
{}\vdots&0&\ddots&\ \ \ \ddots\ \ \ &-1&0
\\
{}0&\vdots&\ddots&-1&2&- 1+{\it {\lambda}}
\\
{}0&0&\ldots&0&-1-{\it {\lambda}}&2
\end {array}
 \right]\,.
 \label{toymo}
 \ee
For this model we revealed, in Ref.~\cite{JMP}, that the simplifying
role of the up-down symmetry need not be decisive. In particular,
although the necessary solutions $E_n$ and $|\psi_n\kt$ of the
underlying time-independent Schr\"{o}dinger's bound-state problem
(\ref{SEb}) remained numerical, we were able to avoid the necessity
of their construction (needed, first of all, in the spectral formula
for the metrics) by the non-numerical construction of the metrics
via the computer-assisted direct solution of the Dieudonn\'{e}'s
equation. In this sense our present text may be read as a
continuation and as a climax of the study initiated in
Ref.~\cite{JMP}.

\section{The energies and metrics}

\subsection{The reality
of the spectra of $H^{(N)}({\lambda})$.}

\begin{figure}[h]                     
\begin{center}                         
\epsfig{file=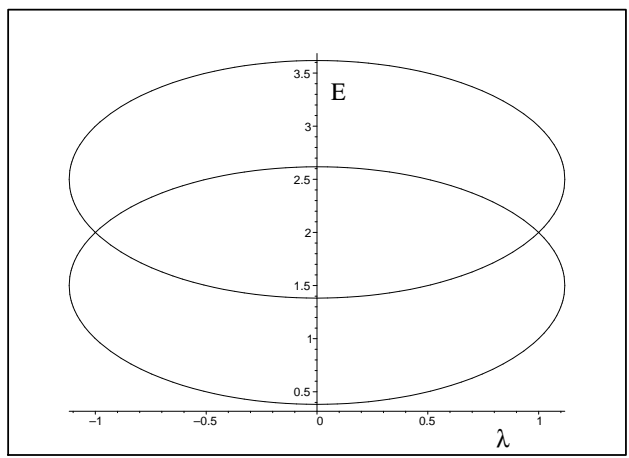,angle=270,width=0.6\textwidth}
\end{center}                         
\vspace{-2mm} \caption{Graphical form of spectrum  at $N=4$.
 \label{firmonej}}
\end{figure}

\begin{figure}[h]                     
\begin{center}                         
\epsfig{file=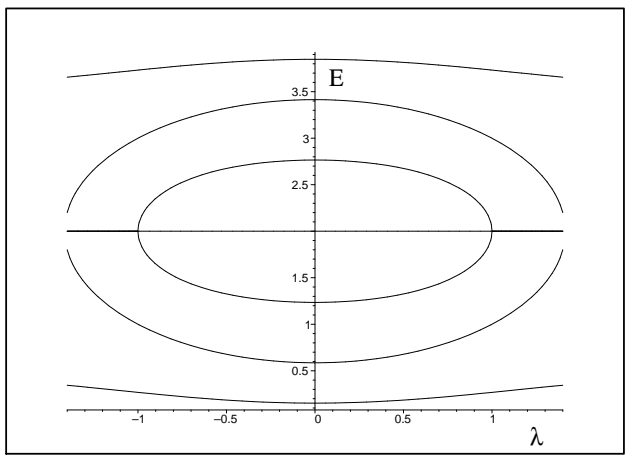,angle=270,width=0.6\textwidth}
\end{center}                         
\vspace{-2mm} \caption{The $\lambda-$dependence of energies at
$N=7$.
 \label{fiststsrmone}}
\end{figure}

At any $N=3,4,\ldots$ and $\lambda \in (-1,1)$ let us consider the
finite-dimensional-matrix $N$ by $N$ toy Hamiltonians
$H^{(N)}({\lambda})$ of Eq.~(\ref{toymo}). They will be shown
suitable for illustration of our present reinterpretation of
Eq.~(\ref{PT}). First of all, the spectrum of their energies is very
easily evaluated at $N=3$,
 $$E_0=2\,,\ \ \ \ \  E_{\pm 1}= 2\pm (2-2\,{{\lambda}}^2)^{1/2}\,$$
as well as at $N=4$,
 $$  E_{\pm 1/2}=3/2\pm 1/2\,(5-4\,{\lambda}^2)^{1/2}\,,\ \ \ \ \ \
 E_{\pm 3/2}=5/2\pm 1/2\,(5-4\,{\lambda}^2)^{1/2}\,$$
(cf. Fig.~\ref{firmonej}) or at $N=5$,
 $$
 E_0=2\,,\ \ \ \ \  E_{\pm 1}= 2\pm (1-{\lambda}^2)^{1/2}
 \,,\ \ \ \ \  E_{\pm 2}= 2\pm (3-{\lambda}^2)^{1/2}\,
 $$
etc. One always encounters precisely four fragile levels which
intersect at $\lambda=\pm 1$ at even $N$ and which do not intersect
at odd $N$. A clear distinction emerges between the even and odd
dimensions $N$: one always finds a $\lambda-$independent central
level $E_0=2$ in the latter case (cf. the $N=7$ illustrative example
in Fig.~\ref{fiststsrmone}). Inside the interval of $\lambda \in
(-1,1)$ the spectrum is discrete, up-down symmetric, non-degenerate
and real at any $N$ (cf. the proof in \cite{JMP}).

%
%
%

\subsection{Dieudonn\'e equation  and its
definite/indefinite solutions. }

Equation (\ref{PT}) may be treated as a linear algebraic constraint
 \be
 \sum_{k=1}^N\,
 \left [
      \left (H^\dagger\right )_{jk}\,\Theta_{kn}
      -\Theta_{jk}\,H_{kn}\right ] =0
 \,,\ \ \ \ \ j,n=1,2,\ldots,N\,, \ \ \ N \leq \infty
   \,
 \label{htot}
 \ee
which is imposed either upon the heavily non-unique positive
definite ansatz
 \be
   \Theta= \left (\Theta^{(N)}_{(M)}\right )^{(S)}=
   \sum_{k=1}^M\, {\mu_k}\,\Theta^{(N)}_{(k)} > 0\,
 \label{e777h}
   \ee
containing the suitable sparse, $(2k-1)-$diagonal components
$\Theta^{(N)}_{(k)}$ and defining the physical Hilbert-space metrics
at any $M \leq N$, or upon the indefinite though still invertible
metric ${\cal P}$ in Krein space ${\bf K}$,
 \be
   \Theta = {\cal P}= \left (\Theta^{(N)}_{(M)}\right )^{(Krein)}=
   \sum_{k=1}^M\, {\nu_k}\,\Theta^{(N)}_{(k)}\,
 \label{e777k}
   \ee
or upon the metric in the slightly more general auxiliary
Pontryagin space $\tilde{\bf K}$,
 \be
   \Theta = {\cal P}= \left (\Theta^{(N)}_{(M)}\right )^{(Pontryagin)}=
   \sum_{k=1}^M\, {\rho_k}\,\Theta^{(N)}_{(k)} \,.
 \label{e777p}
   \ee
In the case of the former Eq.~(\ref{e777k}) we shall assume that the
respective numbers $N_-$ and $N_+$ of  the negative and positive
eigenvalues of $\Theta$ will be roughly the same and, in any case,
infinite in the limit $N \to \infty$. In the latter case of
Eq.~(\ref{e777p}) one of the {\em non-equal} numbers $N_-$ and $N_+$
should stay, by the definition of Pontryagin spaces, finite in the
limit $N \to \infty$.

\section{The
Hilbert/Krein/Pontryagin classification of auxiliary spaces }

For the sake of brevity of our present considerations we shall keep
the dimension $N$ finite and fixed and even. We shall speak about
the ``Krein-space-simulating case" if $N_-=N_+$ and about the
``Pontryagin-space case" if $0 \neq N_-\neq N_+ \neq 0$.

\subsection{The non-numerical diagonal solution of Dieudonn\'e equation.}

In Ref.~\cite{JMP} we may find the explicit and, up to an
inessential overall factor, unique diagonal solution of
Eq.~(\ref{htot}),
 \ben
 \Theta^{(N)}({\lambda})=\left (\Theta^{(N)}_{(1)}\right )^{(S)}=
  \left[ \begin {array}{ccccc}
 \alpha &0&\ldots&0&0
 \\{}0&1&0&\ldots&0\\
 {}\vdots&\ddots&\ddots&\ddots&\vdots
 \\{}0&\ldots&0&1&0
 \\{}0&0&\ldots&0&\alpha
 \end {array} \right]\,
 \label{diago}
 \een
with
 \ben
 \ \ \ \ \alpha=\alpha(\lambda)={\frac
 {1-{\it {\lambda}}}{1+{\it {\lambda}}}}\,.
 \een
Inside the open interval of $\lambda \in (-1,1)$ the latter quantity
remains positive so that just the Hilbert-space solution
(\ref{e777h}) is obtained at $M=1$. In our present paper we may
speak about the $N_-=0$ case or about the signature denoted,  at an
illustrative $N=N_+=8$, by the symbol \fbox{$++++++++$} which
displays the set of the signs of the eigenvalues of $\Theta$ in
question.


\subsection{The  bidiagonal, indefinite solutions of
Dieudonn\'e'\ equation}

Obviously, in a search for the indefinite metrics in ${\bf K}$ or
$\tilde{\bf K}$ we must study  metrics  (\ref{e777k}) or
(\ref{e777p}) with $M=2$ at least. Firstly, recalling the results of
Ref.~\cite{JMP} we find the bidiagonal solution of Eq.~(\ref{htot})
(with $\nu_1=0$ and inessential $\nu_2$ in Eq.~ (\ref{e777k})) which
is very sparse and, up to the overall factor $\nu_2>0$, unique,
 \be
  \Theta^{(N)}({\lambda})=\left (\Theta^{(N)}_{(2)}\right )^{(Krein)}=
 \left[ \begin {array}{cccccc}
  0&\beta
 &0&0&\ldots&0\\{}
 \beta&0
 &1&0&\ldots&0\\{}0&1&\ddots&\ddots&\ddots&\vdots
 \\{}0&\ddots&\ddots&0&1&0
 \\{}\vdots&\ddots&0&1&0 &\beta
 \\{}0&\ldots&0&0&\beta&0
 \end {array} \right]\,.
 \label{bidia}
 \ee
This matrix contains the single positive variable $
\beta=\beta(\lambda)={1-{\it {\lambda}}}$ and may be classified as a
(slightly non-standard) Krein-space metric. Its non-standard status
results from its non-involutivity property
 \ben
 \left [\left (\Theta^{(N)}_{(2)}\right )^{(Krein)} \right ]^2
 \neq I\,.
 \een
This implies that the involutive Krein-space metric ${\cal P}$ must
be constructed via the preliminary diagonalization of matrix
(\ref{bidia}), i.e., via the evaluation of the set of its
eigenvalues and, if needed, eigenvectors. In an illustrative
example, let us pick up $N=8$. Then, at $\beta=1$ the numerically
evaluated sample of eigenvalues is
                      $ \pm 1.87938524$,
                  $\pm 1.53208889$,
                  $\pm 1$ and
                 $\pm 0.347296355$.
In other words, the eight-dimensional matrix (\ref{bidia}) may be
assigned the Krein-space-representing signature (i.e., the set of
the signs of the eigenvalues) \fbox{$++++----$} as well as a certain
``generalized-parity" status at $\beta=1$.

A slightly uncomfortable shortcoming of the similar {\em
constructive} definitions of the auxiliary space ${\bf K}$ lies in
the natural dimension- and Hamiltonian-dependence of the metrics $
\Theta^{(N)}({\lambda})$. Fortunately, their signature is changing
rarely. For example, at the same dimension $N=8$ the transition to a
very small quantity $\beta=0.02$ still gives the spectrum $ \pm
1.80196162$, $ \pm 1.24709165$, $ \pm 0.445529554$, $ \pm
0.00039952$ compatible with the same Krein-space signature
\fbox{$++++----$}.


\begin{table}[t]
\caption{The sample of the coupling-dependence of eigenvalues and of
the signatures for the tridiagonal metrics $\Theta$ of
Eq.~(\ref{trojdia}) at $N=8$.} \label{pexp4}

\vspace{2mm}

\centering
\begin{tabular}{||c|c|c|c||}
\hline \hline
   & {\rm(doubly degenerate)} && \\
  $\lambda$ & {\rm eigenvalues of $\Theta^{(8)}(\lambda)$} &{\rm signature}&{\rm classification } \\
 \hline \hline
 -1&  1.885199025, 1.103209260,  & \fbox{$++++++-  -$}& {\rm acceptable ${\cal P}$,}
 \\
 & 0.1798728781,
 -0.6682811631&  {\rm regular,} &  {\rm Pontryagin space}
 \\
 \hline
 -0.5&  2.325566538, 1.335120809,  & \fbox{$++++++-  -$}& {\rm acceptable ${\cal P}$,}
 \\
 & 0.07897024957,
 -0.9396575972&  {\rm regular,} &  {\rm Pontryagin space}
 \\
 \hline
 0& 2.53208889,  1.34729636,
 & \fbox{$++++0 \ 0-  -$} &  {\rm exceptional,}\\
 &0,
                 -0.879385241& {\rm singular,} & {\rm not acceptable}\\
                 \hline
 0.5&  2.10869763, 0.981936410,  & \fbox{$++++++-  -$}& {\rm acceptable ${\cal P}$,}
 \\
 & 0.0376599205,
 -0.328293960&  {\rm regular,} &  {\rm Pontryagin space}
 \\
 \hline
 0.95&  1.66131164, 0.785034968, & \fbox{$++++++-  -$}& {\rm acceptable ${\cal P}$,}
 \\
 & 0.0809096799,
 -0.0016321045&  {\rm regular,} &  {\rm Pontryagin space}
 \\
 \hline
 1&  1.62348980,
 0.777479066, & \fbox{$++++++0\ 0$}&  {\rm exceptional,}
 \\
 & 0.099031132,
 0&  {\rm singular,} &   {\rm not acceptable}
 \\
 \hline
 \hline
 1.1&  1.555271054,
 0.7692404001, & \fbox{$++++++--$}& {\rm off domain,}
 \\
 & 0.1312085969,
 -0.003231363463&  {\rm regular,} &  {\rm Pontryagin space}
 \\
 \hline
 2&  1.235525737,
 0.8538808152, & \fbox{$++++++--$}& {\rm off domain,}
 \\
 &  0.1508170150,
 -0.04022356696&  {\rm regular,} &  {\rm Pontryagin space}
 \\
 \hline
 200&  1.000040269,
 0.9999844811, & \fbox{$++++++--$}& {\rm far off domain,}
 \\
 &  0.00002499875,
 -0.000024748775&  {\rm close to singular,} &  {\rm Pontryagin space}
 \\
 \hline
 -200&  1.000040631,
 0.9999846191, & \fbox{$++++++--$}& {\rm far off domain,}
 \\
 &  0.00002499875,
 -0.000025248725&  {\rm close to singular,} &  {\rm Pontryagin space}
 \\
 \hline \hline
\end{tabular}
\end{table}

\subsection{The  more-diagonal indefinite solutions of
Dieudonn\'e equation}

According to Ref.~\cite{JMP} the general more-than-two-diagonal and
maximally sparse solution of the Dieudonn\'e equation still has the
unique form at all $N$,
 \be
  \Theta^{(N)}_{(k)}({\lambda})=
 \left[ \begin {array}{ccccccc}
                    \ldots&0&0&z&0&0&\ldots
  \\{}\ldots&0&v&0&v&0&\ldots
  \\{}
 {\large \bf _. } \cdot {\large \bf ^{^.}}
 &v&0&1&0&v&\ddots
  \\{}
 {\large \bf _. } \cdot {\large \bf ^{^.}}
 &0&1&0&1&0&\ddots
  \\{}
 {\large \bf _. } \cdot {\large \bf ^{^.}}
 &1&0&1&0&
 1&\ddots
  \\{}
 {\large \bf _. } \cdot {\large \bf ^{^.}}
 &0&1&0&1&0&\ddots
 \\{}
 {\large \bf _. } \cdot {\large \bf ^{^.}}
 &
 {\large \bf _. } \cdot {\large \bf ^{^.}}
 &
 {\large \bf _. } \cdot {\large \bf ^{^.}}&\vdots&\ddots&\ddots&\ddots
\end {array} \right]\,.
 \label{endiago}
 \ee
It is defined in terms of the two functions of coupling $\lambda$,
 \ben
  z=\gamma({\lambda})
 ={\frac {1-{\it {\lambda}}}{1+{\it {\lambda}}^2}}
 \,,\ \ \ \
  v=\delta({\lambda})={\frac {1}{1+{\it {\lambda}}^2}}\,.
 \label{defi}
  \een
For illustrative purposes we selected just the simplest, tridiagonal
special case,
 \be
 \Theta^{(N)}_{(3)}({\lambda})=
 \left[ \begin {array}{ccccccc}
  0&0 &z&0&\ldots&&0\\
 {}0&v &0&v&0&\ldots&0\\
 {}z&0&1&\ddots&\ddots&\ddots&\vdots
 \\
 {}0&v&\ddots&\ddots&\ddots&v&0
 \\
 {}\vdots&\ddots&\ddots&\ddots&1&0&z
 \\
 {}0&\ldots&0&v&0&v &0
 \\
 {}0&&\ldots&0&z&0&0
 \end {array} \right]
 \label{trojdia}
 \ee
and calculated, numerically, its eigenvalues and signatures at
$N=8$. The sample of results is displayed in Table~\ref{pexp4}. It
shows that in our toy model there emerges the fully general but
still sufficiently elementary and viable  third, Pontryagin-space
alternative to the most usual Hilbert-space signature
\fbox{$++++++++$} with $N_-N_+=0$ and to the PTSQM-related
Krein-space signature \fbox{$++++----$} with $N_--N_+=0$. We have
arrived at our conclusions.


%
%
%

\begin{conj}
In place of the standard PTSQM strategy of choosing the sufficiently
elementary Krein-space pseudometric ${\cal P}$ of Eq.~(\ref{PT}) in
advance, it may prove more efficient to start from {\em any given}
Hamiltonian with real spectrum and to construct  the suitable PTSQM
pseudometric ${\cal P}$ as a diagonalized form of {\em any} sparse
solution $\Theta$ of the Dieudonn\'{e}'s Eq.~(\ref{htot}).
\end{conj}

 \begin{cor}
Once a given Hamiltonian $\hat{H}$ and a self-adjoint and a
boundedly-invertible bounded operator $\Theta$ satisfy
Eq.~(\ref{htot}) with $N \leq \infty$, the latter operator may play
the role of the metric either (i.e., if positive definite) in the
standard Hilbert space ${\bf H}^{(S)}$ of states, or  (i.e., if
$N_-=N_+$) in the auxiliary PTSQM Krein space ${\bf K}$ or, thirdly
(i.e., if $N_-\neq N_+$), in an alternative auxiliary Pontryagin
space $\tilde{\bf K}$.
 \end{cor}

\section{Summary}

Within the present extended version of ${\cal PT}-$symmetric quantum
mechanics we still start working with Hamiltonians $\hat{H}$ defined
in an auxiliary, first, {\em unphysical} Hilbert space ${\bf
H}^{(F)}$ where $\hat{H}\neq \hat{H}^\dagger$ is allowed
non-Hermitian. In an intermediate step we propose to replace ${\bf
H}^{(F)}$ {\em either} by Krein  space ${\bf K}$ {\em or} by
Pontryagin space $\tilde{\bf K}$. Secondly, in place of the
traditional selection of the (indefinite) metric ${\cal P}={\cal
P}^\dagger$  {\em in advance} (i.e., typically, in the form of an
operator of parity) we propose to proceed {\em constructively}. This
means that we only assume that the Hamiltonian is self-adjoint in
${\bf K}$ or $\tilde{\bf K}$, i.e., that it satisfies the
Dieudonn\'{e} equation $\hat{H}^\dagger\,{\cal P}={\cal P}\,\hat{H}$
where the operator (or matrix) ${\cal P}$ is {\em not} given in
advance and must be constructed and/or chosen out of a broader menu.

It has been multiply tested in the past that with many pre-selected
unphysical Hilbert spaces ${\bf H}^{(F)}$ and  Krein-space metrics
${\cal P}$ the validity of the Dieudonn\'{e}'s equation opened the
way towards the necessary and ultimate transition to the ``second",
{\em physical} Hilbert space ${\bf H}^{(S)}$. In our present paper
we tested and verified the feasibility of the similar transition
${\bf H}^{(F)}\to {\bf H}^{(S)}$ under the assumption that the
auxiliary indefinite metric (or, if you wish, the pseudometric)
${\cal P}$ only specifies the less usual, Pontryagin intermediate
space $\tilde{\bf K}$.


\subsection*{Acknowledgment}

Work supported by the GA\v{C}R grant Nr. P203/11/1433, by the
M\v{S}MT ``Doppler Institute" project Nr. LC06002 and by the Inst.
Res. Plan AV0Z10480505.


\end{document}